\documentclass[nofootinbib,prd,twocolumn,superscriptaddress]{revtex4}

\usepackage{amsmath,empheq}
\usepackage{amsfonts}
\usepackage{amsthm}
\usepackage{amssymb}
\usepackage{graphicx}
\usepackage{hyperref}
\usepackage{soul}
\usepackage[normalem]{ulem}
\usepackage{color}

	\newcommand{\ncd}{\newcommand}
	\ncd{\mrm}    {\mathrm}
	\ncd{\beq} {\begin{equation}}
	\ncd{\eeq} {\end{equation}}

	\def\d{{\rm d}}

\graphicspath{{Graphics/}}

\begin{document}
\title{Thermodynamic optimization of a Penrose process: an engineers' approach to black hole thermodynamics}
\date{\today}

\author{A. Bravetti}
\email{bravetti@correo.nucleares.unam.mx}
\address{Instituto de Ciencias Nucleares, Universidad Nacional Aut\'onoma de M\'exico,\\ AP 70543, M\'exico, 
DF 04510, Mexico.}

\author{C. Gruber}
\email{christine.gruber@correo.nucleares.unam.mx}
\address{Instituto de Ciencias Nucleares, Universidad Nacional Aut\'onoma de M\'exico,\\ AP 70543, M\'exico, 
DF 04510, Mexico.}
\address{Institut f\"ur Physik, Universit\"at Oldenburg, D-26111 Oldenburg, Germany}

\author{C. S. Lopez-Monsalvo}
\email{cesar.slm@correo.azc.uam.mx}
\address{CONACYT Research Fellow, Departamento de Energ\'ia, Universidad Aut\'onoma Metropolitana Azcapotzalco, CP 02200, M\'exico, DF, Mexico}

\begin{abstract}
In this work we present a new view on the thermodynamics of black holes introducing effects of 
irreversibility by employing thermodynamic optimization and finite-time thermodynamics. These 
questions are of importance both in physics and in engineering, combining standard thermodynamics 
with optimal control theory in order to find optimal protocols and bounds for realistic processes 
without assuming anything about the microphysics involved. We find general bounds on the maximum work and 
the efficiency of thermodynamic processes involving black holes that can be derived exclusively from the 
knowledge of thermodynamic relations at equilibrium. Since these new bounds consider the finite duration of 
the processes, they are more realistic and stringent than their reversible counterparts. 
To illustrate our arguments, we consider in detail the thermodynamic optimization of a Penrose process, 
i.e.~the problem of finding the least dissipative process extracting all the angular momentum from a 
Kerr black hole in finite time. We discuss the relevance of our results for real 
astrophysical phenomena, for the comparison with laboratory black holes analogues 
and for other theoretical aspects of black hole thermodynamics.
\end{abstract}

\maketitle

\section{Introduction}

Black hole thermodynamics links the classical concepts of gravity with thermodynamic properties of 
gravitational objects such as black holes, and is supposedly determined by the quantum nature of spacetime. 
It is a very interdisciplinary field of research relevant to many areas of investigation, reaching from the 
understanding of astrophysical phenomena such as the production of jets and the origin of active galactic 
nuclei, to more theoretical issues like the AdS/CFT correspondence and the disputed emergent nature of gravity. 

In the context of black hole thermodynamics, many different properties and scenarios have been studied. Recently, 
black holes have been considered from an  ``engineering'' point of view -- including studies on their stability 
with respect to fluctuations and phase transitions \cite{1978Davi,kaburaki1993thermodynamic}, on the construction 
of heat engines involving black holes \cite{kaburaki1991kerr,xi2009reversible,opatrny2012black,johnson2014holographic}, 
on the possibility to extract rotational energy from a black hole 
\cite{penrose1969gravitational,penrose1971extraction,wald1974energy,thorne1974disk,bhat1985energetics,davies1978thermodynamics} 
and on various issues about generalized versions of the second law of thermodynamics and related entropy bounds 
\cite{bekenstein1974generalized,2002Bous}. 
In particular, the extraction of energy from a rotating black hole has been conceptualized as a thought 
experiment known as the Penrose process. In its original formulation, an object is imagined to enter the 
ergosphere of a Kerr black hole and split into two parts. One of the parts falls into the black hole following 
a geodesic along which it has negative energy as seen by an observer at infinity, and the other escapes to 
infinity with positive energy. By energy conservation, the part escaping to infinity possesses an energy greater 
than the original object, therefore energy has been extracted from the hole. Hawking's area theorem guarantees 
that in general the area does not decrease in such process, as must be the case in any classical process 
\cite{davies1978thermodynamics}. When considering also quantum processes, the area can in principle decrease,
but we will not consider any such processes here.

Since this scenario could in principle serve as a model for astrophysical phenomena such as the powering of 
active galactic nuclei, X-ray binaries and quasars \cite{bhat1985energetics}, it is a relevant task to understand 
the maximum efficiency of any such process. In the first models of a Penrose process it was shown that there is 
an upper bound of about $29\%$ on the efficiency \cite{wald1974energy,thorne1974disk}, implying that the resulting 
energy of the outgoing particles is greater than the energy of the original infalling particles, but not as much 
as would be needed for astrophysically reasonable applications. For this reason several extensions have been 
proposed. Piran et al. \cite{piran1975high,piran1977upper} were the first to suggest scattering and annihilation 
processes near the horizon, which became known as the \emph{collisional Penrose process}, however, with still 
modest energy gain. More recently, the interest in energy extraction from black holes has been reignited after 
the proposal by Ba\~nados et al. \cite{banados2009kerr} that black holes surrounded by cold dark matter relics
could act as particle accelerators with collisions at arbitrarily high energies (see also 
\cite{berti2009comment,jacobson2010spinning,wei2010charged,banados2011emergent}). These results were extended 
by Schnittman \cite{schnittman2014revised,schnittman2015distribution}, who proved that  -- by carefully tuning the 
initial conditions for the collision -- an efficiency as high as $1300\%$ can be reached. Berti et al. 
\cite{berti2014ultra} then considered an improved Schnittman's method with pre-collisions in order to obtain 
an energy gain with no theoretical limits and dubbed such mechanism a \emph{super-Penrose process}. 
All these approaches can be considered as \emph{mechanical Penrose processes}, as they employ the notion of 
individual particles falling into the black hole. The resulting efficiency depends on the assumptions about 
the particular trajectories the particles are following and is defined as the ratio between the energy of the 
outgoing particles with respect to their initial energy \cite{lasota2014extracting} {(see also \cite{leiderschneider2015maximal}
and the references therein for a critical discussion on the real
efficiencies and feasibility of such processes)}. 

Non-mechanical Penrose processes have also been considered. Blandford and Znajek \cite{blandford1977electromagnetic} 
suggested a model of electromagnetic extraction of a black hole's rotational energy, known as the \emph{BZ process}. 
Recently \cite{lasota2014extracting} this proposal has been generalized to include energy extraction by means of 
arbitrary fields or matter in what has been called a \emph{generalized Penrose process}. 
A further possibility was put forward by Unruh and Wald, who employed the thermodynamic properties of black holes 
in order to mine the acceleration radiation near the horizon and proved that this can be performed at a huge energy 
rate \cite{unruh1983mine}. This approach to energy extraction from a black hole could be termed 
\emph{thermodynamic Penrose process}. However, the laws of black holes thermodynamics imply that not all the {energy}
extracted from the black hole can be converted into useful work -- we must account also for the exchange of heat and for dissipative effects. 
In this context, a different approach has been recently suggested in \cite{dolan2012pdv,2011CQGra..28w5017D}, where 
limits on the thermodynamic efficiency of Penrose processes for various types of black holes have been derived from 
the first law of black hole thermodynamics alone. Interestingly, these limits precisely coincide with the earlier 
results of an efficiency of $29\%$ for the mechanical Penrose process.  However, the two efficiencies are derived 
from different perspectives and therefore there is no reason why they should coincide. In the mechanical approach, 
the efficiency is defined as the ratio between the energy of the outgoing particle and the initial energy of the 
infalling particles. In the thermodynamic case, the efficiency is defined as the ratio between the work extracted 
and the initial energy (i.e.~mass) of the black hole. Therefore, since the initial energy of the black hole is 
much larger than the initial energy of the particles, an efficiency of $29\%$ from the thermodynamic picture could 
in principle be able to accelerate the particles to arbitrarily high energies, resulting in a very large mechanical 
efficiency. 

The thermodynamic efficiency for black hole processes has been analyzed so far by means of the first law only. 
Consequently, this approach only considers reversible thermodynamic effects and is limited to equilibrium 
thermodynamics. In engineering literature, this would be called the \emph{first law efficiency}, or the 
\emph{energy efficiency} \cite{rosen2002clarifying,kanoglu2007understanding}. In reality however, classical 
equilibrium thermodynamic analysis of processes is not sufficient to describe real processes. In contrast to 
reversible processes, which proceed without losses at an infinitely slow speed \cite{hoffmann2003optimal}, real 
processes are always carried out in finite times, and thus usually have efficiencies that are far from the 
predicted reversible value. One could naively expect that these discrepancies could be overcome by technological 
progress or better implementations of the process. However, it is possible to define general bounds on the 
efficiency of irreversible processes which are inevitable in any finite-time process, and cannot be circumvented. 
To give an example, we consider the efficiency of a Carnot heat engine, which is the largest efficiency possible 
for heat engines, 
	\beq\label{etaCARNOT}
	\eta_{1}^{\rm Carnot}=1- \frac{T_{C}}{T_{H}} \,. 
	\eeq
A Carnot engine is a reversible engine, which means that it operates on infinite time scales and therefore has 
zero power output, i.e. 
	\beq\label{power}
	  \frac{\d W}{\d t} = 0 \,. 
	\eeq
It was shown by Curzon and Alborn \cite{curzon1975efficiency} under the assumption of a finite time scale of 
operation that a realistic Carnot heat engine working at maximum power has the actual efficiency (see also 
\cite{van2005thermodynamic})
	\beq\label{etaCA}
	\eta_{1}^{\rm CA}=1-\sqrt{\frac{T_{C}}{T_{H}}} \,. 
	\eeq
Therefore, the solution of finding optimal processes and realistic bounds on the efficiency depends on the 
particular objective function that is extremized, i.e.~on the physical condition of interest that should be 
optimized. The derivation of such general bounds for irreversible processes and the quest for optimal protocols 
is the principal aim of the interdisciplinary research area comprising engineering, physics and control theory, 
which is called \emph{finite-time thermodynamics} 
\cite{andresen1977thermodynamics,andresen1984thermodynamics,hoffmann1989measures,salamon1998geometry,
salamon2001principles,andresen2011current,andresen2015metrics}, also known as \emph{thermodynamic optimization}
or \emph{entropy generation minimization} in the engineering literature \cite{bejan1996entropy,hoffmann2003optimal}.

Finite-time thermodynamics sets bounds on the efficiency in the same way as classical reversible thermodynamics 
does, but including a further physical constraint, i.e.~that real processes must happen in a finite interval of 
time and with finite resources, and therefore can never be ideally reversible. By carrying out processes in finite 
time, it is not possible to establish perfect equilibrium in each infinitesimal step along the way, and dissipative 
losses accumulate during the process, effectively decreasing the efficiency. The thermodynamic efficiency of real 
systems (e.g.~heat engines) is thus usually smaller than the reversible predictions (e.g.~Carnot's efficiency) and 
so the reversible treatment is in general not very useful for application purposes.  
Therefore, more appropriate measures of efficiency for real thermodynamic processes have been introduced based 
on the second law and on a quantity known as the exergy, whose changes measure the maximum (reversible) work that can be 
extracted from a system in contact with a reservoir. Efficiencies defined on these grounds are usually referred 
to as \emph{second law efficiencies} or \emph{exergy efficiencies} \cite{rosen2002clarifying,kanoglu2007understanding}. 
Using the exergy in the efficiency analysis is more appropriate than the energy in real processes since it permits us
not only to determine the extents of losses, but also their causes and locations, and therefore is a tool to 
optimize the system overall.
Finite-time thermodynamics has been applied in a wide range of fields, both theoretical and practical (see 
e.g.~\cite{andresen2011current,andresen2015metrics,bejan1996entropy,CrooksPRL2012} and references therein), but 
has to the best of our knowledge never been considered in the context of black hole thermodynamics. In our analyses, 
we will use these methods to compute the maximum work that can be obtained from a thermodynamic Penrose process in 
a finite amount of time and obtain general bounds on the possible efficiencies of this mechanism from an energetic 
point of view. 

We will start by reviewing a central principle in finite-time thermodynamics, known as the 
\emph{horse-carrot theorem}, substituting the idealized picture of a reversible process with a discrete sequence 
of thermal equilibria with a heat bath. We will then introduce the notion of thermodynamic length of a process, 
defined via a differential geometric picture of thermodynamics as e.g. in the works of Weinhold and Ruppeiner 
\cite{wein1975,rupp1979,rupp1995}, in order to quantify the dissipative losses along the process, and proceed 
in analogy to \cite{dolan2012pdv,2011CQGra..28w5017D} in order to derive the maximum work that can be extracted 
from a Kerr black hole by a Penrose process from a thermodynamic perspective. In contrast 
to \cite{dolan2012pdv,2011CQGra..28w5017D}, we will consider the black hole in contact (but not in equilibrium) 
with a reservoir, and take into account the dissipated energy along the process. 

Our results thus link two aspects of thermodynamics that are rather distinct in essence: the highly theoretical 
area of black hole thermodynamics, which is still subject of many fundamental debates (though we are starting to 
obtain some experimental access to it by means of black hole analogues \cite{unruh1981experimental,rousseaux2008observation,
belgiorno2010hawking,weinfurtner2011measurement,2015arXiv151103900L}), and the question of non-perfect thermodynamic 
processes, mostly applied in industrial and engineering contexts so far,  which bears important consequences for 
experiments. With this work, we provide a first 
step towards a more realistic and experimental view on black hole thermodynamics, and we expect that our findings 
will lead us to further understanding of the astrophysical nature of black holes, which is largely determined 
by its thermodynamic properties. Further investigations of thermodynamic Penrose processes and their astrophysical 
signatures are to follow, and we hope that with this article we can show the importance of finite-time effects 
on black hole thermodynamics and highlight its potential for future investigations. 

This article is structured as follows. In Section~\ref{sec:FTTD} we will review the basic tools of 
finite-time thermodynamics by introducing the horse-carrot theorem, which is employed to find bounds on the 
dissipation along real processes. In Section~\ref{sec:BHbounds} the thermodynamic geometry of Kerr black holes 
will be presented, and a specific process for the energy extraction selected and worked out in detail. 
Ultimately, we will address the questions of the maximum work and efficiency in Section~\ref{sec:efficiency}, 
and conclude our work.

\section{Finite time thermodynamics and the horse-carrot theorem}\label{sec:FTTD}

Finite-time thermodynamics was initiated in the mid $70$s with the purpose of finding limits on the 
efficiency of thermoynamic and chemical processes carried out in a finite interval of time and to 
revise existing bounds such as Carnot's from reversible thermodynamics, which are based on the 
condition that an infinite amount of time is at hand to carry out the process. This is an unrealistic 
assumption unlikely to hold in practical situations, and finite-time effects are expected to decrease 
the efficiency that can be obtained in a process (see e.g.~\cite{andresen2011current} for a recent 
introductory review). 

One of the central results of finite-time thermodynamics is the derivation of such a revised bound on the 
efficiency from the notion of thermodynamic length, which can be defined in the framework of geometric 
thermodynamics, as e.g.~introduced by Weinhold and Ruppeiner \cite{wein1975,rupp1979}. In these frameworks, 
a metric is defined in the phase space spanned by the thermodynamic state variables describing the system, 
and thus the meaning of distance in this phase space can be investigated. 
The thermodynamic metric of Weinhold is defined as the Hessian of the internal energy of the system with 
respect to the extensive parameters, whereas Ruppeiner's metric employs the entropy of the system as the 
defining potential instead of the internal energy. Stability requirements of thermodynamics require the 
Hessian matrix of the internal energy, $U_{ij}$, to be positive definite, or vice versa, the Hessian of 
the entropy, $S_{ij}$, to be negative definite.  
Therefore, using standard notation with subindices to indicate partial derivatives and Einstein summation 
convention over repeated indices, a well-defined concept of length can be defined for Weinhold's metric 
\cite{wein1975} as 
	\beq\label{TDLW}
	L_{U}=\int_{\gamma}\sqrt{U_{ij}\d {x^{i}} \d {x^{j}}}\,,
	\eeq
where the $x^{i}$ are the extensive state variables of a thermodynamic system, including the entropy, and $\gamma$ 
is the specified path which represents the thermodynamic process under consideration in the abstract space 
spanned by the $x^{i}$. With this definition, it is possible to associate a real number to every thermodynamic 
process between two states of a given system, i.e.~the \emph{thermodynamic length} of such process. 

In analogy, a slightly different notion of thermodynamic length can be introduced using Ruppeiner's metric \cite{rupp1979}, 
given by minus the Hessian of the entropy with respect to the extensive quantities, $S_{ij}$. Since from stability 
considerations the Hessian $S_{ij}$ of the entropy with respect to the extensive variables -- which now include 
the energy instead of the entropy -- is negative definite, Ruppeiner's thermodynamic length can be defined as 
	\beq\label{TDLR}
	L_{S}=\int_{\gamma}\sqrt{-S_{ij}\d {x^{i}} \d {x^{j}}}\,.
	\eeq
The geometries of Weinhold and Ruppeiner and in general the geometric picture of thermodynamics has been studied 
extensively in the literature and therefore we refer the interested reader to some of the main works on the subject 
\cite{weinhold2009classical,rupp1995,rupp2010,rupp2012,mrugala1,MNSS1990,rajeev2008hamilton,CONTACTHAMTD}. Our 
interest here lies in the application of these concepts to finite-time thermodynamics. 
To see this, let us introduce the quantity 
	\beq\label{availability}
	A=U-T_{0}S+p_{0}V\,,
	\eeq
which is called the \emph{availability} or \emph{exergy} of the system \cite{hoffmann1989measures}. Here subscript 
zero refers to intensities associated with the environment. The change in $A$ from an initial state to a final state 
for the system equals (minus) the maximum amount of work that can be extracted (reversibly) from a system that is in 
contact with a thermal reservoir, that is \cite{LL}
	\beq\label{maxWever}
	W^{\rm max}=-\Delta A=-\Delta U+T_{0}\Delta S-p_{0}\Delta V\,.
	\eeq
Note that the relevant feature in order to be able to extract work from the system is that it cannot be in equilibrium with 
the environment, since otherwise $\Delta A=0$. Moreover, if the process is not performed reversibly, part of the 
available work is lost to dissipation and the net work of the process is less than \eqref{maxWever}. 
Consequently, we can define the quantity $(\Delta A)_{\rm dest}$, i.e. the amount of availability that is destroyed during 
the process due to irreversibility. As it turns out \cite{SalamonBerryPRL}, the square of the thermodynamic length 
$L_{U}$ provides a lower bound to this exergy loss of the system, 
	\beq\label{SB1}
	(\Delta A)_{\rm dest}\geq L_{U}^{2}\,\frac{\epsilon}{\tau}\,.
	\eeq
Here, $\epsilon$ is a mean relaxation time that depends on the system and $\tau$ is the total duration of the 
process \cite{SalamonBerryPRL,andresen2011current}. Implicitly assumed in the derivation of this bound is that 
$\epsilon \ll \tau$. For a process carried out in $N$ steps from an equilibrium state to another, we can substitute 
the expression $\epsilon/\tau$ for $1/2N$. 
Therefore the thermodynamic length $L_{U}$ is a measure of dissipation along irreversible processes, also called 
the \emph{price of haste} in the finite-time literature. In view of \eqref{SB1}, we obtain the maximum work that 
can be extracted from a system in contact with a reservoir as a function of the duration $\tau$ of the process as 
	\beq\label{maxWeverfinite}
		W^{\rm max}(\tau) 
		=-\Delta U+T_{0}\Delta S-p_{0}\Delta V-L_{U}^{2}\,\frac{\epsilon}{\tau}\,.
	\eeq

The bound given by \eqref{SB1} is important since it includes information about the irreversibility of the 
process but does not need to assume any particular microscopic model for the irreversible phenomena. The limit 
it sets on the efficiency is more stringent than the standard thermodynamic limit, which predicts 
	\beq\label{revdissipated}
	(\Delta A)_{\rm dest}\geq 0\,, 
	\eeq
but the standard result is recovered if the protocol is allowed to operate in an infinite time. 
In fact, when the total duration of the process diverges, the right hand side of \eqref{SB1} vanishes, thus 
precluding any dissipative losses during the process. 

Equality in \eqref{SB1} is achieved when the \emph{thermodynamic speed} $\d L_{U}/\d t$ of the process is constant, 
thus giving an explicit condition for the best protocol, provided the path is fixed. 
Finally, since the bound is given by the square of the thermodynamic length, if only the initial and final states 
are specified while the path is undefined, machinery from Riemannian geometry dictates that the \emph{shortest paths} 
are represented by the metric's \emph{geodesics}, and therefore the least dissipating protocols can be achieved by 
following geodesic paths (given the duration of the process and the relaxation time). 

In analogy, also Ruppeiner's metric can be employed to define a thermodynamic length $L_{S}$ and a similar bound, 
	\beq\label{SB2}
	(\Delta S)_{\rm irr}\geq L_{S}^{2}\,\frac{\epsilon}{\tau}\,. 
	\eeq
Here, $(\Delta S)_{\rm irr}$ represents the irreversible entropy production of the system and the environment during the 
process. 

Such bounds are derived from scenarios which go under the name of \emph{horse-carrot theorems}, where the horse 
illustrates the system that is being driven over the specified path of the process by successive contact with 
different environments, represented by the carrot \cite{salamon1998geometry}. 

Early applications of the bounds \eqref{SB1} and \eqref{SB2} have been the optimization of the industrial 
distillation process and chemical reactions. Also applications in economics and information coding have been 
found (see \cite{andresen2011current} for a complete review). Moreover, such results have been recently extended 
to mesoscopic systems through the use of fluctuation theorems and the Fisher information metric \cite{CrooksPRL2012}. 
Also investigations of the geodesics for different thermodynamic models have been performed, both in connection 
with phase transitions \cite{brody2009information,PhysRevE.86.051117} and in order to find optimal thermodynamic 
protocols \cite{CrooksPRE2012,CrooksPRL2012,rotskoff2015dynamic}. 

We have now at hand completely general bounds on the efficiency of thermodynamic processes, which do not 
depend on the microscopic details of irreversibility in the system, but only on the thermodynamic properties 
in equilibrium and the time span of the process. Furthermore, at least in principle, they permit us to find the 
optimum {(least dissipative)} protocols. 

As mentioned before, the prototypical and for astrophysical reasons most interesting black hole is the Kerr 
black hole. Commonly considered scenarios involving Kerr black holes as sources of energy are variations of 
the Penrose process, i.e.~the extraction of energy from a Kerr black hole by decreasing its angular momentum 
in different ways. Subsequently, we will consider Penrose processes from a thermodynamic point of view, with 
the aim to compute the maximum amount of work that can be extracted from such processes in a finite interval 
of time, focussing on the least dissipative processes possible for our system, which are the \emph{geodesic paths} 
in the thermodynamic geometry of the Kerr black hole.

\section{Thermodynamic geometry for Kerr black holes and optimal Penrose process}\label{sec:BHbounds}

Since black hole thermodynamics lacks a microscopic description, methods from geometric thermodynamics have been
used extensively in the literature in order to gain some insights about its microscopic nature. In particular, 
analyses of thermodynamic stability and phase transitions for various black holes have been performed using 
different thermodynamic metrics and their scalar curvature (see e.g.~\cite{aaman2003geometry,arcioni2005stability,
ruppeiner2007stability,2007Aman,shen2007thermodynamic,ruppeiner2008thermodynamic,2008Alva,LiuLu,banerjee2011second,
2013Bravc,iofra,Hendi201442,suresh2014thermodynamics,tharanath2014thermodynamic,garcia2014geometric,zhang2015phase,
PhysRevD.92.044013,mansoori2015hessian,aaman2015thermodynamic,WeiPRLinsight}). 
Here we will focus on the Kerr black hole only. First we will review its thermodynamic stability following 
\cite{ruppeiner2007stability} and using results from \cite{aaman2015thermodynamic}. Then we will study the 
geodesics of this thermodynamic geometry, and use them to find the thermodynamic optimum for a Penrose process. 

The entropy $S$ of a Kerr black hole is related to its mass $M$ and its angular momentum $J$ by the relation
	\begin{equation}\label{Kentropy}
	  S(M,J) = 2M^2\left(1+\sqrt{1-\frac{J^{2}}{M^{4}}}\right)\,,
	\end{equation}
where here and in the following we will use Planck units, unless otherwise specified. 
The corresponding thermodynamic geometries of Weinhold and Ruppeiner are defined on a two-dimensional parameter space 
with coordinates $(S,J)$ or $(M,J)$, respectively. 

Before proceeding to calculate the geodesics for these metrics, we shall consider the thermodynamic stability of 
this system. As thermodynamic stability corresponds to positivity of the corresponding Weinhold and Ruppeiner's 
metrics, we can compute one of the two metrics and infer regions of stability and instability from its properties. 
We can use either metric, noting that the two geometries are completely analogous since they are related by a 
conformal factor given by the inverse of the temperature \cite{SalamonRW}. The thermodynamic geometry of the Kerr 
black hole has been studied in detail in the literature, see \cite{ruppeiner2007stability} for a general overview, 
or \cite{aaman2015thermodynamic} for a more in-depth analysis. According to \cite{aaman2015thermodynamic}, it turns 
out that it is possible to transform the coordinates in the thermodynamic manifold from $(M,J)$ to $(t,x)$ such 
that Ruppeiner's line element in parameter space takes the form 
	\beq\label{RKerrtx}
	\d s^{2}_R=\frac{1}{T}\left(\d x^{2}-\d t^{2}\right)\,,
	\eeq
where the temperature defined by 
	\beq\label{KerrTtx}
	T(t,x)=\frac{(t^{2}-x^{2}-2tx)(t^{2}-x^{2}+2tx)}{2(t^{2}-x^{2})^{3}}
	\eeq
is the Hawking temperature, and the transformations are given by 
	\beq\label{trafo1}
	M(t,x) = \frac{t^2 - x^2}{4} \mathrm{~~and~~} J(t,x) = \frac{tx \left( t^2 - x^2 \right)^3}
	{4\left( t^2 - x^2 \right)^2}  \,.
	\eeq
Incidentally, in these coordinates it is easy to check that Weinhold's metric is flat, 
	\beq\label{WKerrtx}
	\d s^{2}_W= \d x^{2}-\d t^{2}\,,
	\eeq
which will be useful in order to find the geodesics.

\subsection{Stability regions}
It is clear from  \eqref{RKerrtx} and \eqref{WKerrtx}  that both Ruppeiner and Weinhold's metrics are in 
general not positive definite, which implies that the system is not stable with respect to any general 
thermodynamic perturbation. However, the condition for a process to be thermodynamically stable can be 
given by 
	\beq\label{StabilityCond}
	\left|\frac{\d x}{\d t}\right|>1\,.
	\eeq 
Let us focus from now on on Weinhold's metric, which is the simplest of the two since it is flat in this case.
The condition \eqref{StabilityCond} implies that the squared length of the process, 
	\beq\label{lengthxt}
	L_{U}^2 = \Delta x^2 - \Delta t^2 \,,
	\eeq
is a positive quantity. 
The stable region in parameter space $(t,x)$ thus defines a wedge at any point, akin to the light cone in 
special relativity. All thermodynamic processes are translated into curves in this parameter space and 
therefore a thermodynamic process is stable if and only if the tangent vector of the corresponding curve 
always lies inside the thermodynamically stable wedge, as if it were a ``spacelike'' process in relativity. 
A ``timelike'' process (negative squared-norm) in this context represents an unstable perturbation of the 
system, which eventually would destroy the black hole. In order to avoid that, we must restrict our 
definition of thermodynamic Penrose processes to stable processes only. 
There are further stability criteria for a process which can be imposed to make it physically plausible 
and which we must implement on our definition of a thermodynamic Penrose process, namely that the temperature 
be greater than zero along the whole process and that the generalized second law of black holes be satisfied, 
or its restricted version respectively, the area law theorem, if only classical processes are involved. 

In the following we will consider geodesics of Weinhold's metric as those processes in a given interval of 
time that minimize the dissipated availability according to \eqref{SB1}  when going from one equilibrium state 
to another. Then we find the particular geodesic corresponding to the process that extracts all the angular 
momentum from an extremal Kerr black hole, i.e. the optimum Penrose process that starts with an initial state 
given by an extremal Kerr black hole and ends up with a Schwarzschild black hole.

\subsection{Geodesics}
From the flat Weinhold metric \eqref{WKerrtx} it is straightforward to write down the exact form of the 
geodesics. In $(t,x)$-space they are straight lines, and can be parameterized in general by 
	\beq\label{param}
	t = a x + b \,.
	\eeq
However, the knowledge of the exact form of the geodesics in this space neither gives physical intuition 
about the process nor makes it clear how to fix $a$ and $b$ in order to obtain a process from the extremal Kerr 
to the Schwarzschild black hole as intended. We therefore need to reverse the transformation and return to 
$(M,J)$-space. The transformations \eqref{trafo1} are in general not uniquely invertible -- on the contrary, 
they are highly degenerate. However, using the parameterization \eqref{param} it is possible to narrow down 
the inverse functions $x(M,a,b)$ to two possibilities, and thus obtain the two possible expressions for 
$J(M,a,b)$ from \eqref{trafo1}. This essentially gives the evolution of the geodesic in $(M,J)$-space. 
 We can also express the entropy in terms of $(t,x)$, 
	\beq\label{KerrStx}
	S(t,x)= \frac{\left( t^2 - x^2 \right)^4}{4\left( t^2 + x^2 \right)^2} \,,
	\eeq
and use the parameterization \eqref{param} to give it in terms of $M$, $a$ and $b$. The expressions for 
$S(x,a,b)$ and $T(x,a,b)$ together with the mass $M(x,a,b)$ and the angular momentum $J(x,a,b)$ are then used 
to impose physical conditions on the parameters $a$ and $b$ and thus to pick a single geodesic corresponding 
to the desired optimal Penrose process.

\subsection{Physical boundary conditions and optimal Penrose process}
As already discussed, thermodynamic requirements impose three conditions for a process to be viable. Firstly, 
the thermodynamic length has to be real and positive. Secondly, the area law cannot be violated, i.e.~the 
entropy of the black hole along the process cannot decrease. Thirdly, the temperature should be positive throughout the process. 
Let us consider the \emph{particular} Penrose process which removes all the rotational energy of an extreme Kerr black hole. To do so, we fix the initial and final points of the process in 
$(M,J)$-space to be $p_{i}=(M_i,M_i^{2})$ and $p_{f}=(M_f,0)$. Additionally, we set boundary conditions on the 
temperature. We impose that temperature is zero at the initial point, $T(t_i,x_i) = 0$, and increases towards 
the Schwarzschild limit $T(t_f,x_f) = 1/8M_f$. 

These conditions, together with the expressions of the thermodynamic quantities in terms of $x$, $a$ and $b$, 
allow to completely determine the values for the parameters $a$ and $b$ for this specific process, obtaining
	\beq\label{aandb}
	a = 0 \quad {\rm and} \quad b= \sqrt {2M_{i}\, \left( 1+\sqrt {2} \right) } \,.
	\eeq
Furthermore, we can express the final mass $M_{f}$ as a function of $M_{i}$ as well, 
	\beq
	M_{f}(M_{i})=\frac{M_{i}}{2} \left(1+\sqrt{2}\right) \,.
	\eeq
A puzzling feature of this expression is the fact that the mass increases in this process -- although we 
did not expect the final mass to be the irreducible mass along an irreversible process, it is however quite 
surprising that the final mass is actually \emph{larger} than the initial mass. 

Using these results, we can rewrite all the thermodynamic quantities solely as functions of the mass $M$, with the 
initial mass $M_{i}$ as the only free input parameter. The function $J(M)$ thus obtained is the optimum Penrose 
process extracting all the angular momentum from an extremal black hole with the least dissipation, 
	\beq\label{JM}
	J(M)={\frac {\sqrt {2M_{i}}\sqrt {1+\sqrt {2}}\sqrt {2\,M_{i} \,\sqrt {2}+2\,M_{i}-4\,M}
	{M}^{3}}{ \left( M_{i}\,\sqrt {2}+M_{i}-M \right) ^{2}}} \,.
	\eeq
The length of the process can equally be given in terms of the initial mass as 
	\beq\label{LM}
	L(M_{i})=\sqrt {2 M_{i}(\sqrt {2}-1)} \,.
	\eeq
Here, $M_i$ correctly serves as a mere scaling factor to the results without changing the qualitative evolutions, 
and will be set to unity in the following. 

\begin{figure}[h]
    \centering
    \includegraphics[width=.4\textwidth]{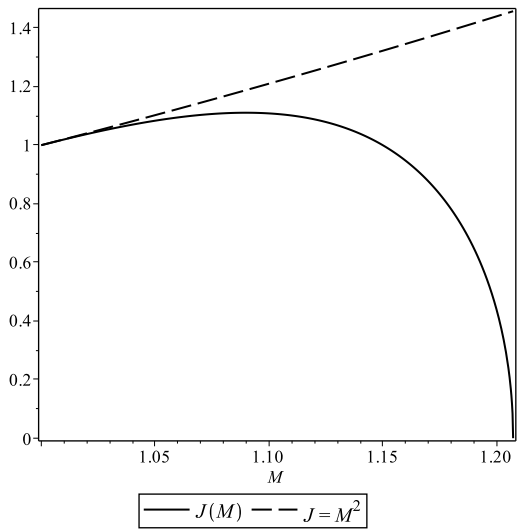}
    \caption{Minimum dissipation, finite-time protocol for stopping a rotating black hole. 
    The curves show the evolution of $J(M)$ along the geodesic path (solid line), compared with the corresponding values for 
    extremal limit $J=M^{2}$ (dashed line). }
    \label{fig:JoverM}
\end{figure}

Figure~\ref{fig:JoverM} shows the evolution of the angular momentum as a function of the mass, together with the 
extremal limit $J = M^2$, i.e. it shows the concrete form of the optimum Penrose process in the (physical) 
$(M,J)$-space. The process correctly starts at the extremal point $J_{i}=M_{i}^{2}=1$ and always stays below 
the extremal limit, until it stops when the Schwarzschild condition $J_{f}=0$ is reached, for the corresponding
value of the mass $M_{f}\simeq 1.207$. As already mentioned before, the final value of the mass is larger than the 
initial one, while the angular momentum at first increases slightly before declining to zero. Thus, in order to 
extract all of the angular momentum, and under the presumption to minimize dissipative losses (which is equivalent 
to following the geodesics), we have to put in more energy than we get out, which leads to the increase of the black 
hole's mass during the process. We will return to this point in the next section, where a detailed analysis of the 
efficiency is performed. 

\begin{figure}[h]
    \centering
    \includegraphics[width=.4\textwidth]{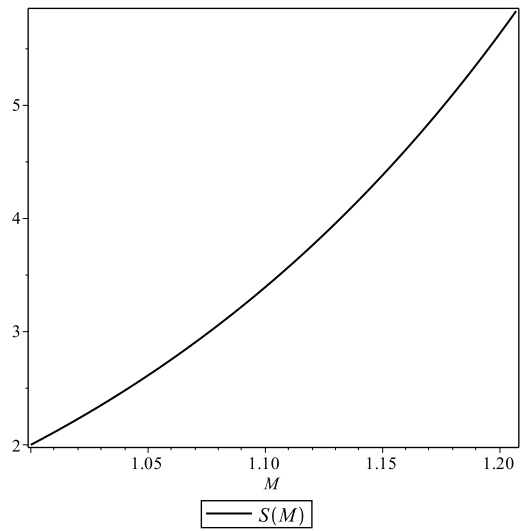}
    \caption{Evolution of the entropy $S$ of the Kerr black hole as a function of the mass $M$ along the geodesic 
    path.}
    \label{fig:entropy}
\end{figure}

Figure~\ref{fig:entropy} shows the evolution of the entropy of the black hole through the optimum process as a 
function of the black hole mass, which is found to be monotonously increasing. This is in accordance with the 
standard area law theorem, meaning that the process is classical. 

\begin{figure}[h]
    \centering
    \includegraphics[width=.4\textwidth]{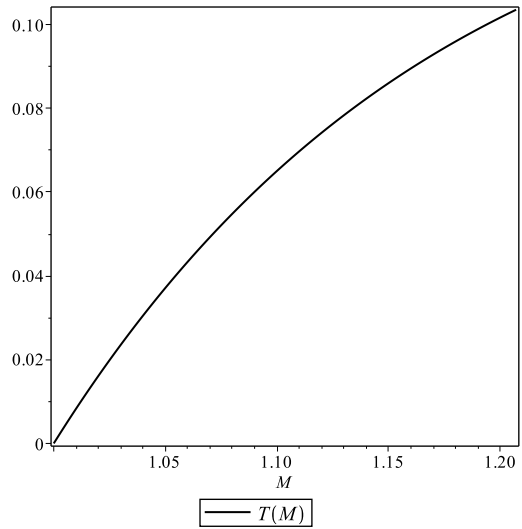}
    \caption{Evolution of the temperature $T$ of the Kerr black hole as a function of the mass $M$ along the 
    geodesic path.}
    \label{fig:temp}
\end{figure}

Figure~\ref{fig:temp} shows the evolution of temperature of the black hole along the optimal process. As expected, 
it increases from zero to its final value $T_f \simeq 0.104$, coinciding with that for a Schwarzschild black hole 
with mass $M_f\simeq 1.207$. Moreover, it is clearly always positive, meaning that all the thermodynamic 
requirements aresatisfied for this process (note that the value $a=0$ guarantees that the process has real 
thermodynamic length and hence is stable).

\section{Maximum work and efficiencies} \label{sec:efficiency} 

In the last section, we have investigated in detail the thermodynamic geometry of Kerr black holes and 
addressed the issues of finding the stability conditions and optimizing the Penrose process that extracts 
all the angular momentum from an extremal black hole. In this section we will compute the work that can 
be extracted from such process, and introduce the notion of exergy efficiency, using the horse-carrot 
theorem \eqref{SB1} to revise the bound on the efficiency of the Penrose process from a thermodynamic perspective. 

Let us start with a brief summary of the arguments in \cite{dolan2012pdv,2011CQGra..28w5017D}. 
We would like to extract energy from a Kerr black hole by decreasing the angular momentum from its extremal 
value down to zero. Following \cite{dolan2012pdv,2011CQGra..28w5017D} we consider the black hole as adiabatically 
isolated. Then the first law of thermodynamics implies that
	\beq\label{1stlawadiabatic}
	W_{\rm adiabatic}=-\Delta M=M_{i}-M_{f} \,,
	\eeq 
i.e.~the work that is extracted during an adiabatic process coincides with the change of energy of the black hole.
Therefore, the maximum amount of work that can be extracted (reversibly) in an adiabatic evolution from an extremal Kerr 
black hole with $J_{i}=M_i^{2}$ to a Schwarzschild black hole with $J_{f}=0$ reduces to the problem of calculating the 
final mass. In the adiabatic case $M_{f}$ is the irreducible mass, 
	\beq\label{Mf}
	 M_f = \frac{M_i}{\sqrt{2}} \,.
	\eeq
Thus, the maximum amount of work that can be extracted from an adiabatic evolution of an extremal Kerr black hole 
down to a Schwarzschild black hole is 
	\beq\label{maxW}
	W^{\rm max}_{\rm adiabatic}=\left(1-\frac{1}{\sqrt{2}}\right)M_{i}\,.
	\eeq
Since the \emph{energy efficiency} is defined as the ratio between the work done by the system and the energy that 
has been supplied, we find that in this case 
	\beq\label{efficiencydef}
	\eta_{1}=\frac{W^{\rm max}_{\rm adiabatic}}{M_i} = 1-\frac{1}{\sqrt{2}} \simeq 0.29 \,.
	\eeq
Therefore in principle one can extract an energy as high as $29\%$ of the initial black hole mass, which for an individual 
nearby particle is a huge amount of energy. However, this calculation does not account for irreversible losses. Moreover, 
the amount of energy that we have thus calculated is the \emph{total} work extracted in the full process from an extremal 
state to a Schwarzschild one. Considering these factors, the work extracted in a realistic situation can be drastically 
smaller than the bound~\eqref{maxW} due to the combined effect of irreversible losses and to the fact that the black hole 
is unlikely to be extremal in its initial configuration. 

In the following,  we will depart from the analysis in \cite{dolan2012pdv,2011CQGra..28w5017D} and consider the black hole 
in contact (but not in equilibrium) with a reservoir. As discussed in the introduction, in this case the maximum amount of 
work that can be extracted (reversibly) from a Kerr black hole is 
\eqref{maxWever}
	\beq\label{maxWeverKerrideal}
	W^{\rm max}_{\rm Kerr}=-\Delta M+T_{0}\Delta S-\Omega_{0}\Delta J\,,
	\eeq
while the optimum work output in a finite amount of time follows from \eqref{maxWeverfinite} and reads
	\beq\label{maxWeverKerr}
	W^{\rm max}_{\rm Kerr}(\tau)=-\Delta M+T_{0}\Delta S-\Omega_{0}\Delta J-L^{2}_{U}\frac{\epsilon}{\tau}\,.
	\eeq 
Note that this value can be significantly greater than \eqref{1stlawadiabatic}, due to the fact that we can now also extract 
energy from the reservoir. Moreover, if the initial and final states of the black hole are fixed, this maximum work is reached 
when the protocol follows a geodesic in the thermodynamic manifold. 

From equation \eqref{maxWeverKerr} and using the equations for the optimal Penrose process derived in the preceding section
(equation \eqref{Kentropy} to compute the initial and final entropy as a function of the mass, \eqref{aandb} for the final mass, 
\eqref{LM} for the thermodynamic length), we can compute the general formula for the maximum amount of work that can be extracted 
from a Kerr black hole in contact with a reservoir at temperature $T_{0}>0$ and $\Omega_{0}=0$, evolving from its extremal state
down to the Schwarzschild state, as 
	\beq\label{maxWeverKerrfinite}
	\begin{split}
	W^{\rm max}_{\rm Kerr}({\tau})=\,&M_{i}\left[(1+2\sqrt{2})T_{0}M_{i}\right.\\
	&~~~~~~~\left.-(\sqrt{2}-1)\left(\frac{1}{2}+2\frac{\epsilon}{\tau}\right)\right]\,.
	\end{split}
	\eeq
Here, the universe is considered as the reservoir, and due to its large scale observed isotropy it is fair to assume $\Omega_{0}=0$. 
Clearly, in equation \eqref{maxWeverKerrfinite} the first term is quadratic in $M_{i}$ and always positive, while the second term 
is linear in $M_{i}$ and always negative. Therefore, depending on the values of $M_{i}$ and $T_{0}$, we will either be able to 
\emph{extract} work from the black hole (in the case that the first term is greater and the second term can be neglected) or we will 
have to \emph{do} work on the black hole in order for the process to occur (if that the second term is greater and the first one can 
be discarded). Moreover, in the case when the factor $T_{0} M_{i}$ is of order $10^{-2}$, we obtain an intermediate situation in which 
both terms are relevant in the calculation of the maximum work. 

In the following we fix $T_{0}$ to be the temperature of the Cosmic Microwave Background (CMB) radiation, i.e.~$T_{0}=3K\simeq 2 \times
10^{-32}\,T_{\rm p}.$ The condition for the process to have positive or negative work production is then given by $M_{i}$ being greater or 
less than $\sim 7.6\times10^{30}\,M_{\rm p}$ respectively (see Figure~\ref{fig:workoutput} below), which lies below the mass of the earth. 
For solar mass black holes with $M_{i}\sim10^{38}\,M_{\rm p}$, the term linear in $M_i$ is thus negligible, and the optimized thermodynamic 
Penrose process derived here gives a total work output of about 
	\beq\label{efficientcase}
	W\sim 10^{44} E_{\rm p} \sim 10^{53} J\,.
	\eeq
This is already a much bigger value than the bound obtained in the adiabatic case -- c.f.~using equation \eqref{maxW}.
Moreover, in contrast to the adiabatic scenario, in this case the energy output is so high that even a fraction of the process can generate 
highly energetic phenomena such as e.g.~gamma ray bursts. The situation is reversed for `small' black holes with initial 
masses below the critical value $7.6\times10^{30} \,M_{\rm p}$. For such black holes we find that the thermodynamic process of extraction 
of energy by removing the angular momentum is inefficient, since work must be provided to the system. Figure~\ref{fig:workoutput} shows 
the total work output for the optimized thermodynamic Penrose process as a function of the initial mass according to \eqref{maxWeverKerrfinite}
and with $T_{0}$ the CMB temperature. Notice that for $0<M_{i}<7.6\times 10^{30}\,M_{\rm p}$ the total work is negative, while for 
$M_{i}>7.6\times 10^{30}\,M_{\rm p}$ it is positive and increasing as the square of the initial mass. 

\begin{figure}[h]
    \centering
    \includegraphics[width=.49\textwidth]{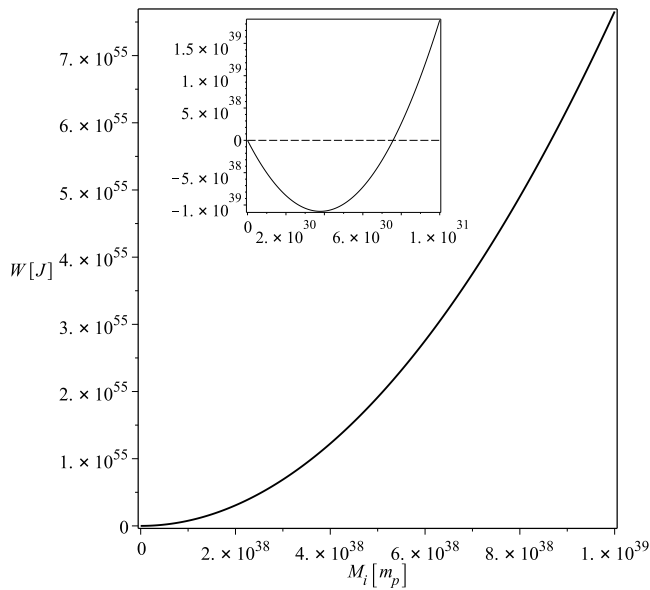} 
    \caption{Total work done by a black hole along the optimal thermodynamic Penrose process as a function of the initial mass. Work is 
    expressed in Joule and mass is in terms of Planck mass. Here we have used $\epsilon/\tau=1/10$, and $T_{0}$ is the CMB temperature.}
    \label{fig:workoutput}
\end{figure}

Let us now focus on the efficiency of the process. For this purpose we restrict ourselves to the cases where the black hole 
\emph{does} work during the process, i.e.~to black holes with initial masses $M_{i}>7.6\times10^{30}\,M_{\rm p}$. 

First we consider the energy efficiency. In this case the total work output is given by \eqref{maxWeverKerrfinite}. 
The total energy input during the process is given by the black hole initial energy -- as in \eqref{efficiencydef} --
plus the energy supplied by the surrounding in the form of heat and work. Since the work term vanishes because $\Omega_{0}=0$, 
we can estimate the total energy supply as 
	\beq
	E_{\rm in}=M_{i}+T_{0}\Delta S\,,
	\eeq 
where $\Delta S$ is the increase in entropy of the black hole during the process. Therefore the energy efficiency reads
	\beq\label{eta1nonisolated}
	\eta_{1}=\frac{W^{\rm max}_{\rm Kerr}(\tau)}{E_{\rm in}}=1-\frac{\sqrt {2} + 4(\sqrt {2}-1)\epsilon/\tau}
	  {2\left[(1+2\sqrt{2})T_{0}M_{i}+1\right]}\,.
	\eeq
For the values $T_{0}=3\,K$ and $\epsilon/\tau=1/10$ used here, we have 
	\begin{align}
	\eta_{1}&=1\%\,\,\, \qquad {\rm for} \qquad M_{i}\sim 8\times 10^{30}\,M_{\rm p}\\
	\eta_{1}&=6\%\,\,\, \qquad {\rm for} \qquad M_{i}\sim  10^{31}\,M_{\rm p}\\
	\eta_{1}&=73\% \qquad {\rm for} \qquad M_{i}\sim 10^{32}\,M_{\rm p}\\
	\eta_{1}&=96\% \qquad {\rm for} \qquad M_{i}\sim 10^{33}\,M_{\rm p}\,,
	\end{align}
meaning that for black holes with $M_{i}\leq 10^{31}\,M_{\rm p}$ only a small part of the total energy input can be converted into 
useful work, while for black holes with $M_{i}> 10^{31}\,M_{\rm p}$ the energy received from the environment is almost completely 
transformed into useful work, with an energy efficiency tending to $1$ as the initial mass increases -- c.f.~equation 
\eqref{eta1nonisolated}. The interpretation of this result is straightforward: for black holes allowed to interact with the universe 
as its reservoir, the process of work extraction is more efficient for larger black holes. 

The energy efficiency $\eta_{1}$ being based only on the first law does not provide a good estimate on how far the process is from 
being optimal. As we have already mentioned in the introduction, such an estimate can be achieved in engineering by means of the 
so-called \emph{exergy efficiency}. There exist several definitions in the literature (see e.g. \cite{kanoglu2007understanding}). 
In our case, we will consider 
	\beq\label{exergyefdef}
	\eta_{2}=\frac{W^{\rm out}}{W^{\rm max}}=\frac{-\Delta A-(\Delta A)_{\rm dest}}{-\Delta A}=
	  1+\frac{(\Delta A)_{\rm dest}}{\Delta A}\,,
	\eeq
where in general $W^{\rm out}$ is the net work output of the process, $W^{\rm max}$ is defined by \eqref{maxWever} and 
$(\Delta A)_{\rm dest}$ is the exergy destroyed due to irreversibilities. Note that $\Delta A$ is negative when work is \emph{done} 
by the system, and that this is exactly the regime in which we are interested, corresponding to $M_{i}>7.6 \,M_{\rm p}$ (see Figure 
\ref{fig:workoutput}). From equation \eqref{exergyefdef} it is then clear that 
	\beq\label{eta2limits}
	0\leq\eta_{2}\leq1
	\eeq 
and that for any reversible process for which $(\Delta A)_{\rm dest}=0$ the exergy efficiency is always $1$. This is of course just 
restating the fact that reversible processes do not have any losses and therefore they all operate at the best possible (ideal) 
exergy efficiency. 

Note further that we apply the definition of $\eta_{2}$ only in the region where the work output along the process is positive, i.e. 
for large black hole masses. For masses below the threshold value of $M_{i} \simeq 7.6 \,M_{\rm p}$, the availability $\Delta A$ 
turns positive, leading to $\eta_{2} > 1$, which means that the definition \eqref{exergyefdef} loses its meaning and should not be 
used in this form. 

Considering realistic processes, we see from the horse-carrot theorem \eqref{SB1} that $\eta_{2}$ is 
maximal along the geodesics connecting the initial and final states, with the maximum value given by
	\beq\label{exergyeffmax}
	\eta_{2}^{\rm max}=1+\frac{L_{U}^{2}}{\Delta A}\,\frac{\epsilon}{\tau}\,. 
	\eeq
where again $L_{U}$ is the length of the process as computed using Weinhold's metric, $\epsilon$ is a mean 
relaxation time that depends on the system and $\tau$ is the total duration of the process. 
For the optimized Penrose process considered here, equation \eqref{exergyeffmax} reads
	\beq\label{eta2Kerr}
	\eta_{2}=1-\frac{4(\sqrt{2}-1)}{1-\sqrt{2}+2(1+2\sqrt{2})T_{0}M_{i}}\frac{\epsilon}{\tau}\,,
	\eeq
which for $T_{0}M_{i}>10^{-2}$ and fixed $\epsilon/\tau$ is between $0$ and $1$ and is an increasing function of $M_{i}$, 
as expected. With the values $T_{0}=3\,K$ and $\epsilon/\tau=1/10$ used here, we have 
	\begin{align}
	\eta_{2}&=16\% \qquad {\rm for} \qquad M_{i}\sim 8\times 10^{30}\,M_{\rm p}\\
	\eta_{2}&=53\% \qquad {\rm for} \qquad M_{i}\sim  10^{31}\,M_{\rm p}\\
	\eta_{2}&=98\% \qquad {\rm for} \qquad M_{i}\sim 10^{32}\,M_{\rm p}\,,
	\end{align}
which confirms that for astrophysical black holes the irreversible losses along the optimized Penrose process considered here 
are practically irrelevant.

From the complete analysis of the maximum work output and of the efficiencies $\eta_{1}$ and $\eta_{2}$ for the optimal thermodynamic 
Penrose process found here, we conclude that for astrophysical black holes with masses $M_{i} > 10^{32}\,M_{\rm p}$ in thermal contact 
with the universe, the work extraction by means of removing the angular momentum is a very powerful and efficient source of energy.

\section{Discussion}

The main aim of thermodynamics is to provide general bounds about maximum work and efficiency, so general that they do not depend on the 
way a practical process is realized. No matter in which way a heat engine is designed and implemented, it is well-known that the 
thermodynamic efficiency can never exceed Carnot's efficiency. In the same spirit, considering the recent attempts to optimize the energy 
extraction from a black hole by means of defining particular processes or trajectories or trying different types of matter, it is of interest 
to find general bounds on the possible extraction of useful energy from a black hole. To achieve results independent of the particular process 
in question, a thermodynamic approach based on the laws of black holes thermodynamics is in order. 
However, equilibrium thermodynamics usually provides too optimistic bounds, which are far from realistic situations, as is the case for 
Carnot's bound on the efficiency of heat engines. The reason is because equilibrium thermodynamics allow for reversible processes, i.e. 
processes taking place in an infinite amount of time. 

In this work we have presented a new approach to study the thermodynamic properties of black holes, considering 
effects from finite-time thermodynamics. We have used results from the horse-carrot theorem, i.e.~equations 
\eqref{SB1} and \eqref{SB2}, providing universal bounds on the efficiency of realistic processes, and requiring only 
the knowledge of the equilibrium fundamental relation of the system. To illustrate the idea, we have considered 
a particular example: a thermodynamic version of the Penrose process for a Kerr black hole, i.e.~a process extracting 
the rotational energy of the black hole without destroying it. Using the bound given by equation \eqref{SB1}, we have 
computed the optimal thermodynamic Penrose process, i.e.~the process that extracts all the angular momentum from a Kerr 
black hole with the least dissipation generated (c.f.~Figure \ref{fig:JoverM}). This optimum process is obtained by a 
geodesic in Weinhold's thermodynamic geometry, demonstrating the usefulness of a geometric approach to thermodynamics. 

Presenting our arguments we have introduced several concepts borrowed from engineering, and calculated and compared 
the efficiencies of the thermodynamic Penrose process as derived from the first and second law. From this combined 
analysis we get a more complete picture of the characteristics of such process and its optimization in different 
regards. 
We believe that concepts such as the availability (or exergy) and the related efficiency will play a major role 
in future discussions about the efficiencies of realistic thermodynamic processes involving black holes. 

From an astrophysical point of view, these analyses are of interest due to the recent attention towards Penrose-like 
processes and their applications in realistic scenarios. 
Our results show that there is a huge amount of energy that can be in principle extracted from a Kerr black hole by 
reducing its angular momentum. If we consider the black hole as an isolated system, the work that can be extracted 
coincides with the difference between the initial and final masses of the black hole and amounts to $29\%$ of the black 
hole initial energy, being linearly proportional to the black hole initial energy -- c.f.~equation \eqref{maxW}. 
During the extraction of work from the black hole, its mass decreases. This result shows that astrophysical black holes 
are capable of doing huge work on systems connected with them. In the case of a solar mass black hole with 
$M_{i}\sim 10^{38}\,M_{\rm p}$ the extracted work reaches $10^{47}\,J$, which is already a reasonable value for making it 
a candidate for a gamma ray burst. However, we need to keep in mind that this amount of energy is the 
total energy that can be extracted when passing from an extremal black hole to a Schwarzschild one. For real situations, 
i.e.~for black holes whose angular momentum is far from extremal and which do not reach the Schwarzschild limit at the 
end of the process, this amount of work should be considerably reduced. 

Considering the case in which the black hole is in contact with an environment, in principle the work that can be extracted 
is much bigger, since energy can be gained from the surroundings. 
A direct consequence of this fact is that the final mass does not necessarily have to be smaller than the initial mass in order 
to have the black hole do work, and thus a much wider range of allowed mechanisms is conceivable. Here, we have computed the maximum 
work that can be extracted from a black hole passing from its extremal state to its Schwarzschild state in a finite amount of time 
-- c.f.~equation \eqref{maxWeverKerrfinite}. The result contains a term proportional to the black hole mass, akin to the case of 
an isolated black hole, but also features a second term proportional to the square of the mass times the CMB temperature, which is 
dominant in the expression for black holes with masses of the order of or larger than $10^{31}\,M_{\rm p}$. 
In particular, for a solar mass black hole the total work output along the optimized protocol we considered is $10^{53}\,J$, much 
larger than in the case of an isolated black hole -- see Figure~\ref{fig:workoutput}. Even though this number is expected to be 
reduced for realistic black holes, which are never really extremal and unlikely to reach the Schwarzschild limit, the value in the 
non-isolated case is so high that this correction should not change the conclusion that this process could in principle generate 
ultra-highly energetic astrophysical phenomena.
It is worth remarking that here we were also including corrections due to irreversibilities along the process, which however 
hardly diminishes the resulting work output. 

Apart from astrophysics, 
a related field of application and possible test bed for our results is the creation of analogue black holes. Since it is possible to 
reproduce black holes and test their properties in the laboratory (see e.g.~\cite{unruh1981experimental,rousseaux2008observation,
belgiorno2010hawking,weinfurtner2011measurement,2015arXiv151103900L}), it would be interesting to compare the experimental results with 
thermodynamic predictions. However, in the context of table-top laboratory experiments, it is even more important to consider realistic 
thermodynamic models including dissipative corrections. Ordinary systems such as cytochrome chains in mitochondria show that the 
inclusion of finite-time modifications can lead to a very good agreement with physical reality \cite{andresen2011current}, and the 
same is expected for analogue gravity. Analogue black holes could thus provide an ideal scenario to test predictions from finite-time 
thermodynamics. 

From a theoretical point of view, the arguments leading to the bound \eqref{exergyeffmax} are independent of the type of black hole and 
process involved, and can therefore be applied to any (stable) thermodynamic process in any black hole for which the fundamental relation 
is known, including cyclic processes like Carnot or Stirling engines. In case of the AdS/CFT correspondence, this means that the scheme 
for deriving bounds on the efficiency of processes involving black holes can be further extended to providing bounds on processes and engines 
in the dual field theories (see e.g.~the discussion in \cite{johnson2014holographic}), and the calculated bounds in the form of \eqref{etaCA} 
or \eqref{exergyeffmax} will be more meaningful than a classical Carnot bound. 
Moreover, the general bounds \eqref{SB1} and \eqref{SB2} are finite-time revisions of the ordinary second law of thermodynamics. Therefore their
study in the context of black holes could provide insights related to the generalized second law, various entropy bounds and the holographic 
principle. 
Ultimately, we would like to mention that the investigations presented here are not limited to ``conventional'' black hole horizons, but could 
in principle be extended to cosmological counterparts such as (A)dS horizons, or Rindler horizons in accelerating reference frames. \\

In summary, we presented the optimization of the thermodynamic Penrose process following from a combined analysis of black hole thermodynamics, 
thermodynamic geometry and control theory and optimization. By connecting these concepts, we have seen how the still rather abstract concepts of 
black hole thermodynamics and the even more abstract concepts of thermodynamic geometry can be successfully employed in the description of 
astrophysically observable highly energetic phenomena. The merits of these results lie in the strong predictive power of thermodynamics.

\section*{Acknowledgements}
AB acknowledges the A. della Riccia Foundation (Florence, Italy) for financial support. CG was supported by 
funding from the DFG Research Training Group 1620 `Models of Gravity' and by an UNAM postdoctoral fellowship 
program. Part of the work of CSLM was funded by an UNAM-DGAPA post-doctoral fellowship. This work has been 
supported by the DGAPA-UNAM Grant No. 113514.

\bibliography{GTD.bib}

\end{document}